\documentstyle[12pt,epsf]{article}
\renewcommand{\baselinestretch}{1.1}
 1
 1
 1

\renewcommand{\thefootnote}{\fnsymbol{footnote}}
\def\lsim{\mathrel{\raise.3ex\hbox{$<$\kern-.75em\lower1ex\hbox{$\sim$}}}}
\def\gsim{\mathrel{\raise.3ex\hbox{$>$\kern-.75em\lower1ex\hbox{$\sim$}}}}

\begin{document}
\begin{titlepage}
\noindent
\thispagestyle{empty}
\renewcommand{\thefootnote}{\fnsymbol{footnote}}

\renewcommand{\baselinestretch}{0.9}
{\small
\begin{flushright}
{\bf TTP96-11}\footnote[1]{The postscript file of this
preprint, including figures, is available via anonymous ftp at
ftp://www-ttp.physik.uni-karlsruhe.de (129.13.102.139) as 
/ttp96-11/ttp96-11.ps 
or via www at http://www-ttp.physik.uni-karlsruhe.de/cgi-bin/preprints/.}
\hfill
{\bf hep-ph/9605311}\\
{\bf MPI/PhT/96-24}
\hfill
{\bf April 1996}\\
{\bf DPT/96/34}
\hfill
\mbox{}
\end{flushright}
}
\renewcommand{\baselinestretch}{1.}

\begin{center}
\begin{Large}
TOP QUARK PRODUCTION TO ${\cal{O}}(\alpha_s^2)$\footnote[2]{
                To appear in the proceedings of the workshop 
                ``Physics with $e^+e^-$ Linear Colliders''.
   }\footnote[3]{
     Work supported by BMFT under Contract 056KA93P6, 
     DFG under Contract Ku502/6-1 and INTAS under Contract INTAS-93-0744.}\\  
\end{Large}
\end{center}

\vspace{0.5cm}

 K.G.~Chetyrkin$^{a}$,
 A.H. Hoang$^{b}$\footnote[4]{Supported by 
              Graduiertenkolleg Elementarteilchenphysik, Karlsruhe.},
 J.H.~K\"uhn$^{b}$, 
 M.~Steinhauser$^{b}$\footnotemark[4],
 T.~Teubner$^{c}$

\vspace{-.6cm}

{\it

\begin{itemize}
\begin{center}
\item[$^a$]
   Max-Planck-Institut f\"ur Physik, Werner-Heisenberg-Institut,\\
   F\"ohringer Ring 6, D-80805 Munich, Germany\\
   and\\
   Institute for Nuclear Research, Russian Academy of Sciences,\\
   60th October Anniversary Prospect 7a, Moscow 117312, Russia
\item[$^b$]
   Institut f\"ur Theoretische Teilchenphysik,\\ 
   Universit\"at Karlsruhe, 
   D-76128 Karlsruhe, Germany\\  
\item[$^c$]
   Department of Physics,\\
   University of Durham, Durham, DH1 3LE, UK
\end{center}
\end{itemize}

}

\vspace{1.0cm}

\addtocounter{footnote}{-4}
\begin{abstract}
\noindent
{\small
The cross-section for the production of $t\bar{t}$ pairs in 
$e^+e^-$-annihilation via a virtual photon is determined
up to order $\alpha_s^2$. For the cm-energies of interest the effect 
of the top mass may not be considered as a small parameter and the full
mass dependence must be included.
}
\end{abstract}
\end{titlepage}

%
%
%
\noindent
Quark mass effects can often be considered as small perturbations in the
analysis of the total cross sections for hadron production in 
$e^+e^-$-annihilation. The approximation  $m_q=0$, supplemented by 
``mass corrections'' of order $m_q^2/s$ is adequate 
for most purposes
and has lead to reliable predictions including terms of order
$\alpha_s^3$~\cite{CheKueKwiRep}. 
The situation is drastically different for top
quark production at a linear collider, where $2\,M_t$ and $E_{cm}=\sqrt{s}$ 
will be of comparable magnitude throughout.
With this motivation in mind the QCD corrections to the vector current
correlator have been calculated in~\cite{HoaKueTeu95,
CheKueSte95,CheHoaKueSteTeu96}
up to ${\cal{O}}(\alpha_s^2)$ including the full quark mass dependence.
Since the corresponding results for the axial vector current are not yet
available, the subsequent discussion is strictly applicable for the
production through the virtual photon only. The results presented
below should therefore not be
considered as absolute predictions but rather as indicative for the
magnitude and importance of the second order corrections. Our notation 
and conventions are based on~\cite{CheKueSte95}. 
The cross-section normalized to the
point cross-section can be cast into the following form:
\begin{eqnarray}
R & = & \frac{\sigma(e^+e^-\to\gamma^*\to t\bar
   t)}{\sigma_{point}}
\nonumber\\[2mm] & = &
  Q_t^2\,\bigg[\,
  R^{(0)} 
  + \bigg(\frac{\alpha_s(M_t^2)}
         {\pi}\bigg)\,C_F\,R^{(1)}
\nonumber\\[2mm] & & \qquad
 + \bigg(\frac{\alpha_s(M_t^2)}{\pi}\bigg)^2\,
 \bigg(\,C_F^2\,R_A^{(2)} + C_F\,C_A\,R_{NA}^{(2)} +
        C_F\,T\,n_l\,R_l^{(2)} + C_F\,T\,R_F^{(2)}
 \,\bigg)
 \,\bigg]
\nonumber\\[2mm] & & 
 + \sum\limits_{q=u,d,s,c,b} Q_q^2\,
 \bigg(\frac{\alpha_s(M_t^2)}
         {\pi}\bigg)^2\,C_F\,T\,R_g^{(2)}
\,.
\end{eqnarray}
$M_t$ denotes the top quark pole mass and $\alpha_s(M_t^2)$ the 
$\overline{\mbox{MS}}$-renormalized strong coupling constant at
the scale $M_t$ with
$n_l=5$ light flavours. The abelian and non-abelian parts $R_A^{(2)}$
and $R_{NA}^{(2)}$ are taken from~\cite{CheKueSte95}, where the 
Pad\'e approximation method has been employed, whereas
$R_l^{(2)}$ and $R_F^{(2)}$
originate from massless and top quark loop insertions, respectively,
into the gluon propagator and are given in closed analytical form
in~\cite{HoaKueTeu95}. 
The last term, $R_g^{(2)}$ originates from gluon splitting into 
$t\bar t$ and has been calculated in~\cite{HoaJezKueTeu94}.
To fix our notation explicitely we note in passing that
\begin{eqnarray}
R^{(0)} & = & \frac{3}{2}\,\beta\,(3-\beta^2)\,,\nonumber\\[2mm]
R^{(1)} & = & 3\,\bigg\{\,
\frac{\left( 3 - {\beta^2} \right) \,\left( 1 + {\beta^2} \right) }{2
    }\,\bigg[\, 2\,\mbox{Li}_2(p) + \mbox{Li}_2({p^2}) + 
     \ln p\,\Big( 2\,\ln(1 - p) + \ln(1 + p) \Big) 
      \,\bigg] \,\nonumber\,\\ 
 & & \mbox{}\quad - 
  \beta\,( 3 - {\beta^2} ) \,
   \Big( 2\,\ln(1 - p) + \ln(1 + p) \Big)  
\nonumber\,\\ 
 & & \mbox{}\quad 
  - \frac{\left( 1 - \beta \right) \,
     \left( 33 - 39\,\beta - 17\,{\beta^2} + 7\,{\beta^3} \right) }{16}\,
   \ln p\,
  + \frac{3\,\beta\,\left( 5 - 3\,{\beta^2} \right) }{8}
\,\bigg\}\,,
\end{eqnarray}
where
\begin{equation}
p \, = \, \frac{1-\beta}{1+\beta}\,,\qquad
\beta \, = \, \sqrt{1-4\,\frac{M_t^2}{s}}
\,.
\end{equation}
$\beta$ is the velocity of one of the produced top quarks in the 
$t\bar{t}$ cm frame.
The result for the different contributions $R_x^{(i)}$ are presented in
Table~\ref{tab} for various energies, adopting a top mass value of
$175$~GeV. The error indicated for $R_A^{(2)}$ and $R_{NA}^{(2)}$
originates from the possible choice of different Pad\'e approximants. As
$R_l^{(2)}$, $R_F^{(2)}$ and $R_g^{(2)}$ are known analytically no error
is specified in this case. $R^{(2)}$ denotes the sum of all second order
correction functions.
\begin{table}[t]
\begin{center}
\begin{tabular}{|c||r|r|r|r|} \hline
$\sqrt{s}$ [GeV] & 400 & 500 & 600 & 700 \\\hline 
$\beta$          & 0.48& 0.71& 0.81& 0.87\\\hline\hline
$R^{(0)}$ & 2.0084 & 2.6673 & 2.8513 & 2.9228  \\ \hline\hline
$C_F\,R^{(1)}$ & 17.6874 & 10.7599 & 7.7942 & 6.2463  \\ \hline\hline
$C_F^2\,R_A^{(2)}$    & 3.5(1)  & -21.4(1)  & -19.63(1)  & -16.32(1) \\ \hline
$C_F\,C_A\,R_{NA}^{(2)}$ & 117.8(2) & 31.9(1)  & 4.03(1)  & -8.61(1) \\ \hline
$n_l C_F\,T\,R_l^{(2)}$  & -32.4356 & -6.8356 & 0.8496 & 4.1789 \\ \hline
$C_F\,T\,R_F^{(2)}$ & 0.7044 & 1.0441 & 1.2375 & 1.3834 \\ \hline
$\sum_q Q_q^2/Q_t^2\,C_F\,T\,R_g^{(2)}$ 
            & $1.2\,\,\, 10^{-6}$ & 0.0005 & 0.0051 & 0.0209 \\ \hline\hline
$R^{(2)}$ & 89.5(3) & 4.7(2) & -13.50(2) & -19.35(2) \\ \hline\hline
$R^{(\alpha_s=0.115)}$ & 1.200 & 1.348 & 1.376 & 1.382 \\ \hline
$R^{(\alpha_s=0.120)}$ & 1.213 & 1.354 & 1.380 & 1.385 \\ \hline
$R^{(\alpha_s=0.125)}$ & 1.227 & 1.360 & 1.384 & 1.388 \\ \hline
\end{tabular}
\end{center}
\caption{ Results for the different contributions to $R$ for
$M_t=175$~GeV and the cm-energies $E_{cm}=400,\,500,\,600,\,700$~GeV.
The total cross-section via photon exchange is given for the
three different values of $\alpha_s(M_t^2)$ corresponding to
$\alpha_s(M_Z^2)=0.115,\,0.120$ and $0.125$.
}
\label{tab}
\end{table}
To arrive at the prediction for $R$, the correction functions have to be
multiplied by the quark charge and 
the corresponding power of the strong coupling
constant evaluated at the scale $M_t$. For 
$\alpha_s(M_Z^2) = 0.115,\,0.120$ and $0.125$ 
this corresponds to 
$\alpha_s(M_t^2) = 0.105,\,0.109$ and $0.113$,
respectively. The complete predictions for the three values of 
$\alpha_s$ are also presented in Table~\ref{tab}.
\begin{figure}[hb]
 \begin{center}
 \begin{tabular}{c}
   \epsfxsize=11.5cm
   \leavevmode
   \epsffile[110 330 460 520]{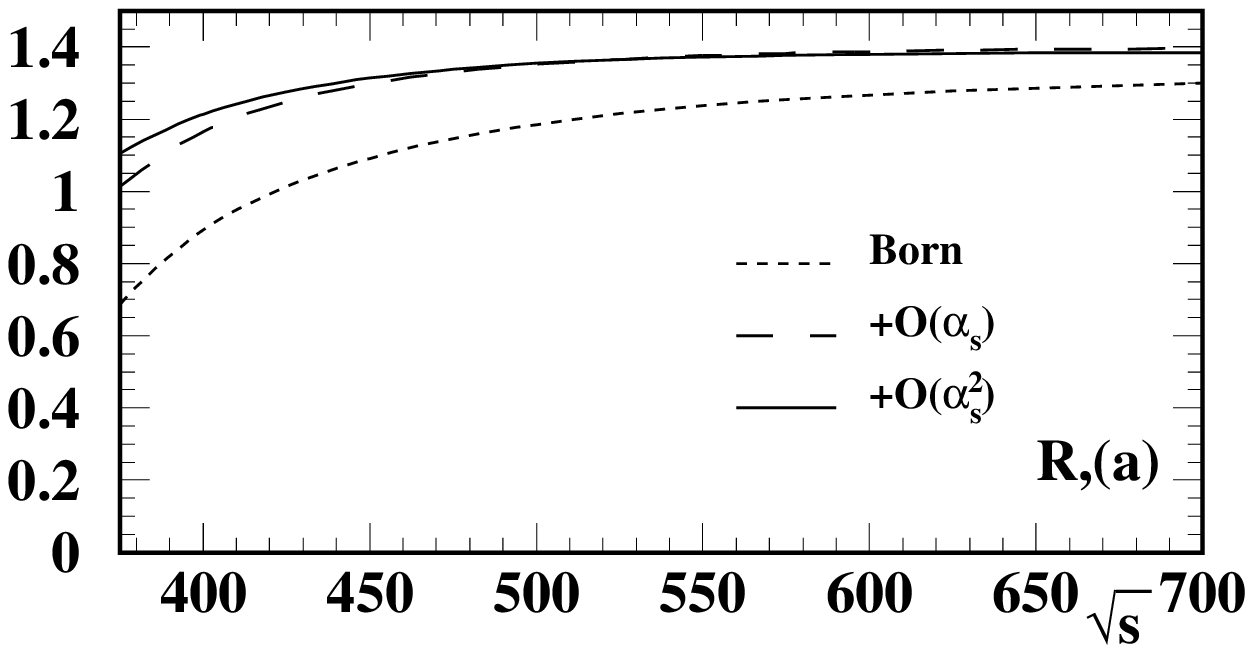}
   \\
   \epsfxsize=11.5cm
   \leavevmode
   \epsffile[110 330 460 520]{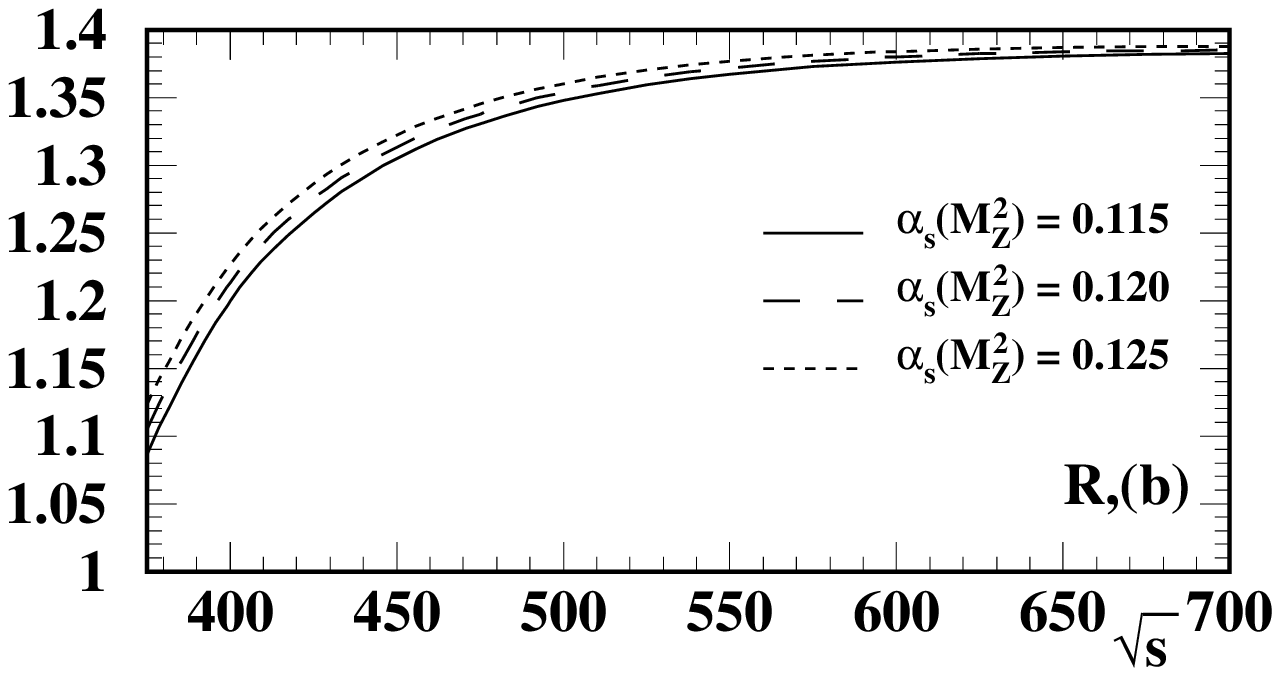}
 \end{tabular}
 \caption{\label{figrfull}$R$ as function of $\protect\sqrt{s}$.
          In figure (b) the ${\cal O}(\alpha_s^2)$ prediction to $R$
          is plotted for different choices of $\alpha_s(M_Z^2)$.
        }
 \end{center}
\end{figure}
In Fig.~\ref{figrfull}a $R$ is plotted against $\sqrt{s}$ including
successively higher orders in $\alpha_s$. 
The dependence on the choice of $\alpha_s(M_Z^2)$ is shown in 
Fig.~\ref{figrfull}b.

\sloppy
\raggedright
\def\app#1#2#3{{\it Act. Phys. Pol. }{\bf B #1} (#2) #3}
\def\apa#1#2#3{{\it Act. Phys. Austr.}{\bf #1} (#2) #3}
\def\lhc{Proc. LHC Workshop, CERN 90-10}
\def\npb#1#2#3{{\it Nucl. Phys. }{\bf B #1} (#2) #3}
\def\plb#1#2#3{{\it Phys. Lett. }{\bf B #1} (#2) #3}
\def\prd#1#2#3{{\it Phys. Rev. }{\bf D #1} (#2) #3}
\def\pR#1#2#3{{\it Phys. Rev. }{\bf #1} (#2) #3}
\def\prl#1#2#3{{\it Phys. Rev. Lett. }{\bf #1} (#2) #3}
\def\prc#1#2#3{{\it Phys. Reports }{\bf #1} (#2) #3}
\def\cpc#1#2#3{{\it Comp. Phys. Commun. }{\bf #1} (#2) #3}
\def\nim#1#2#3{{\it Nucl. Inst. Meth. }{\bf #1} (#2) #3}
\def\pr#1#2#3{{\it Phys. Reports }{\bf #1} (#2) #3}
\def\sovnp#1#2#3{{\it Sov. J. Nucl. Phys. }{\bf #1} (#2) #3}
\def\jl#1#2#3{{\it JETP Lett. }{\bf #1} (#2) #3}
\def\jet#1#2#3{{\it JETP Lett. }{\bf #1} (#2) #3}
\def\zpc#1#2#3{{\it Z. Phys. }{\bf C #1} (#2) #3}
\def\ptp#1#2#3{{\it Prog.~Theor.~Phys.~}{\bf #1} (#2) #3}
\def\nca#1#2#3{{\it Nouvo~Cim.~}{\bf #1A} (#2) #3}

\end{document}